\documentclass[10pt]{article}
\usepackage[left=2cm, right=2cm, top=2cm, bottom=2cm]{geometry}

\usepackage{booktabs}
\usepackage{cite}
\usepackage{siunitx}
\usepackage[colorlinks,linkcolor=black,anchorcolor=black,citecolor=blue]{hyperref}
\usepackage{amsmath, amsfonts, amssymb} 

\usepackage{lastpage,fancyhdr,graphicx}
\usepackage{epstopdf}
\usepackage{color}

\definecolor{Gray}{gray}{.25}

\usepackage{graphicx}

\usepackage{sidecap}

\usepackage{wrapfig}
\usepackage[pscoord]{eso-pic}
\usepackage[fulladjust]{marginnote}
\reversemarginpar

\begin{document}
\vspace*{0.35in}

\begin{flushleft}
{\Large
\textbf\newline{Stochastic Frequency Fluctuation Super-Resolution Imaging}
}
\newline
\\
Yifan Chen\textsuperscript{1},
Chieh Tsao\textsuperscript{1.2},
Hendrik Utzat\textsuperscript{1,2*}
\\
\bigskip
\textsuperscript{\textit{1}} Department of Chemistry, University of California, Berkeley, California 94720, USA
\\
\textsuperscript{\textit{2}} Materials Science Division, Lawrence Berkeley National Lab, Berkeley, California 94720, USA
\\
* \href{mailto:hutzat@berkeley.edu}{hutzat@berkeley.edu}

\end{flushleft}


\section*{Abstract} 
The inherent non-linearity of intensity correlation functions can be used to spatially distinguish identical emitters beyond the diffraction limit, as achieved, for example, in Super-Resolution Optical Fluctuation Imaging (SOFI). Here, we propose a complementary concept based on spectral correlation functions, termed Spectral Fluctuation Super-Resolution (SFSR) imaging. Through theoretical and computational analysis, we show that spatially resolving time-frequency correlation functions in the image plane can improve the imaging resolution by a factor of $\sqrt2$ in most cases and up to twofold for strictly two emitters. This improvement is achieved by quantifying the degree of correlation in spectral fluctuations across the spatial domain. Experimentally, SFSR can be implemented using a combination of interferometry and photon-correlation measurements. The method works for non-blinking emitters and stochastic spectral fluctuations with arbitrary temporal statistics. This suggests its utility in super-resolution microscopy of quantum emitters at low temperatures, where spectral diffusion is often more pronounced than emitter blinking.

\section{Introduction}
The Abbe -- or diffraction -- limit implies that two identical emitters spaced closer than about half of the emission wavelength cannot be distinguished in classical far-field fluorescence microscopy \cite{abbe1873beitrage}. As a result, images of sets of point-like emitters are obscured by the spatial convolution with an instrument-specific point spread function (PSF).

To overcome the diffraction limit, different super-resolution techniques have been introduced. Stochastic Optical Reconstruction Microscopy (STORM) \cite{rust2006sub} and Photoactivated Localization Microscopy (PALM) \cite{betzig2006imaging} localize emitters with high-precision of a few nanometers based on their temporal distinguishability from independent stochastic photoblinking. Stimulated Emission Depletion (STED) microscopy \cite{hell1994breaking} employs two laser pulses: the first excites a set of fluorophores, and the second, shaped like a donut, depletes a peripheral sub-set around the focal spot center via stimulated emission. This selective deactivation leaves only a small central region fluorescing, thereby beating the Abbe limit for the excitation beam. However, the near-saturation laser excitation intensity in STED can photobleach or heat-damage the sample.

\begin{figure}[htbp]
\centering\includegraphics[width=10cm]{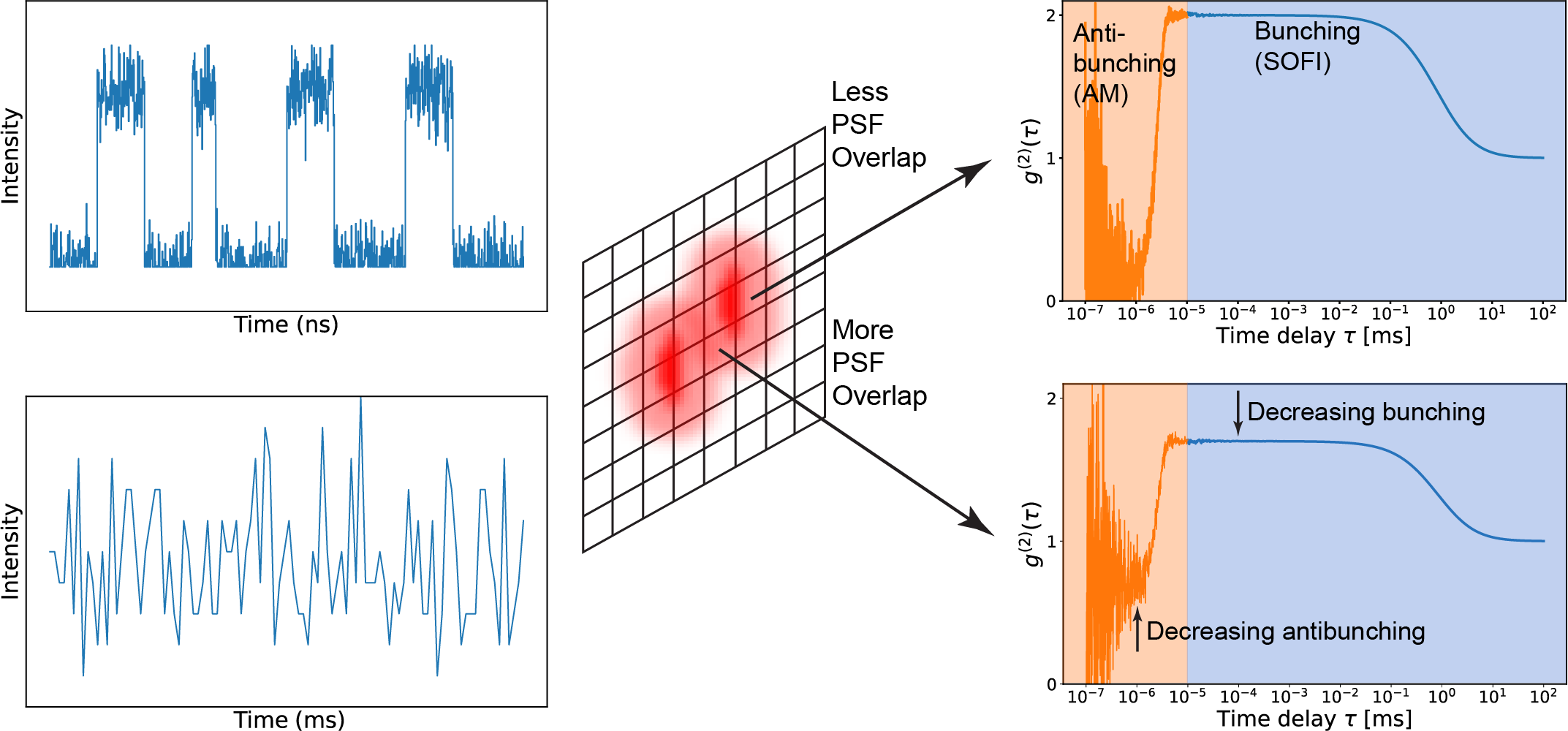}
\caption{Depiction of the ICM concept. ICM relies on stochastic fluorescence fluctuations on different timescales and reconstructs fluorescence images based on temporal photon correlation measurements, typically using the second-order correlation function ($g^{(2)}(\tau)$). Schematic fluorescence intensity traces at two selected pixels of two independent emitters in the image plane are shown. For pixels receiving photons from only one emitter (minimal PSF overlap), maximal antibunching and bunching is observed on timescales of the spontaneous emission (nanoseconds), and intensity fluctuations (nano-milliseconds), respectively. In contrast, pixels receiving photons from both emitters show reduced antibunching and bunching.}
\label{Fig.1}
\end{figure}

Methods in Intensity Correlation Microscopy (ICM) use the non-linearity of temporal photon correlation functions to achieve a narrowed effective PSF. Super-Resolution Optical Fluctuation Imaging (SOFI) \cite{dertinger2009fast} and Antibunching Microscopy (AM) \cite{PhysRevA.85.033812,schwartz2013super} both use the second-order intensity-correlation function $g^{(2)}(\tau)$, where $\tau$ is the time delay between two photon detection events (Fig. \ref{Fig.1}). SOFI relies on emitter intensity fluctuations leading to bunched photon statistics and an elevated $g^{(2)}(\tau)$ with a decay on the characteristic fluctuation timescale, typically microseconds to milliseconds. AM relies on photon anti-bunching, for example produced by single-photon emitters. The corresponding $g^{(2)}(\tau)$ exhibit a dip with a width defined by the emission lifetime, typically nanoseconds, and which indicates the inability of simultaneous emission of two photons. Both SOFI and AM spatially resolve $g^{(2)}(\tau)$ in the image plane e.g. using multi-pixel single-photon detectors. For a given pixel in the image plane, the degree of correlation -- bunching in SOFI and anti-bunching in AM -- depends on the relative overlap of the PSFs of different emitters. Pixels in the region of small PSF overlap show stronger bunching for longer $\tau$ and stronger antibunching on shorter $\tau$. Pixels with large PSF overlap display weaker bunching and antibunching. The spatial variation of the degree of correlation, which is inherently nonlinear, reduces the width of the effective PSF. For example, spatially resolving the second-order photon correlation function reduces the effective PSF in width by a factor of $\sqrt{2}$, which is leveraged as resolution advantage in image reconstruction.

SOFI and AM have different utility profiles. AM requires emitters with photon anti-bunching and faces challenges with slow image acquisition compared to SOFI because the rate of photon coincidence detection for typical single emitter brightness is smaller for short $\tau$ around nanoseconds than $\tau$ between microseconds to milliseconds.  

Compared to STORM and PALM microscopy, SOFI shows reduced acquisition times \cite{dertinger2009fast}.  Relatively lower excitation powers in SOFI further minimize the risk of photo-bleaching and heating compared to STED. Comparatively easier instrumentation renders the correlation approach also practically advantageous in many cases, motivating its extension.

In this concept paper, we introduce Spectral-Fluctuation Super-Resolution (SFSR) imaging, which represents a new way of achieving super-resolution based on stochastic emitter frequency fluctuations. We explore spatially-resolved spectral correlation functions as the image contrast, leveraging same advantages of SOFI as a frequency analog with complementary utility, for example the localization of emitters without any intensity fluctuations, such as high-quality quantum emitters at low temperatures with minimal ensemble inhomogeneous broadening. Like intensity correlation imaging, spectral correlation imaging provides a $\sqrt{2}$-fold resolution advantage for spectral correlation to second order. SFSR is agnostic to the specific temporal stochastics of frequency fluctuations. While SFRS can in principle reconstruct images based on pico- to nanoseconds fluctuations, more typical scenarios will use slower timescale spectral diffusion dynamics from hundreds of nanoseconds to milliseconds \cite{beyler2013direct,Coolen:09}. As a result, SFRS, like SOFI, has a potentially better signal-to-noise ratio (SNR) compared to AM.

\section{Concept and Theory}
\subsection{Experimental concept of SFSR imaging}
\begin{figure}[htbp]
\centering\includegraphics[width=10cm]{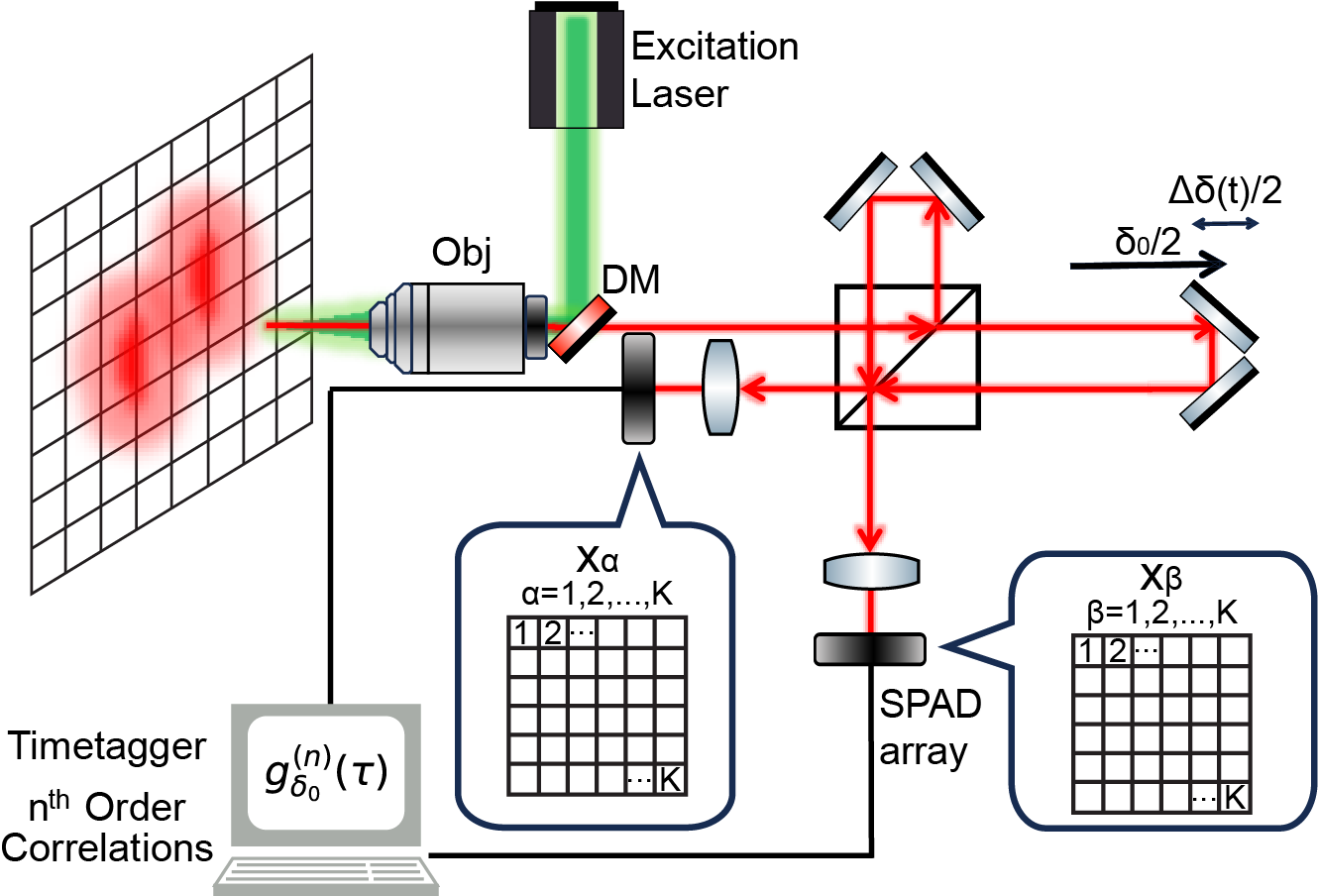}
\caption{One experimental implementation of SFSR imaging. Obj refers to the objective lens and DM stands for the dichroic mirror. The collected emission is modulated by the interferometer with a time-dependent path-length difference and the pixelwise temporal photon correlation at different $\tau$ is measured with SPAD arrays. Pixel-reassigned images are reconstructed with the acquired spatio-temporal frequency correlations.}
\label{Fig.2}
\end{figure}

We chose to discuss a specific implementation of SFSR imaging based on Photon-Correlation Fourier Spectroscopy (PCFS) \cite{brokmann2006photon,PhysRevLett.100.027403,Coolen:09}. PCFS measures the multi-timescale single-emitter spectral correlations \cite{brokmann2006photon} and is therefore a well-suited platform for spectral correlation imaging. Fig. \ref{Fig.2} shows the integration of PCFS with Wide-Field Fluorescence Microscopy (WFM) to perform SFSR. Emitter scene illumination and emission collection are achieved through an objective lens. Excitation light is removed with a dichroic mirror and the emission is directed into a Michelson interferometer. The interferometer has an adjustable path-length difference $\delta_0$ plus a time-varying modulation thereof $\Delta \delta(t)$ \cite{brokmann2006photon,PhysRevLett.100.027403,Coolen:09}. Spatially-resolved time-tagged single-photon detection is performed with two multi-pixel detectors e.g. Single-Photon Avalanche Diodes (SPAD) arrays in the image planes of the two conjugate output ports of the interferometer. Measuring the temporal photon correlations across a wide range of $\tau$ between different pixels at the two ports and at different $\delta_0$ for the time of the modulation $\Delta \delta(t)$, achieves pixel-reassigned images with the spatio-temporal frequency correlation as contrast function.

\subsection{Theory of PCFS——one-emitter, single-pixel case}
We revisit the theory of PCFS \cite{brokmann2006photon,PhysRevLett.100.027403,Coolen:09} as a starting point for our SFSR introduction. Consider the case of a single emitter with two conjugated single-pixel detectors, denoted $x_\alpha,\alpha=1$ and $x_\beta,\beta=1$. Introduction of a scanning Michelson interferometer before these detectors modulates the intensities detected at the two ports of the interferometer by detectors $x_\alpha$ and $x_\beta$:

\begin{equation}
\label{Eq.1}
I_{x_\alpha}\propto 1+\cos(\delta(t)\cdot\omega(t)/c),\quad I_{x_\beta}\propto 1-\cos(\delta(t)\cdot\omega(t)/c),
\end{equation}
where $\delta(t)=\delta_0+\Delta \delta(t)$ is the time-dependent path-length difference of the stage, and $\omega(t)$ is the time-dependent center emission frequency of the emitter. These modulated intensities are then read out with photon counting electronics, where the second-order temporal photon auto- and cross-correlations ($g_{\delta_0,\parallel}^{(2)}$ and $g_{\delta_0,\times}^{(2)}$) are calculated as
\begin{equation}
\begin{aligned}
&g_{\delta_0,\parallel}^{(2)}(\tau)=\frac{\left<I_{x_\alpha}(t)I_{x_\alpha}(t+\tau)\right>}{\left<I_{x_\alpha}(t)\right>\left<I_{x_\alpha}(t+\tau)\right>},\quad g_{\delta_0,\times}^{(2)}(\tau)=\frac{\left<I_{x_\alpha}(t)I_{x_\beta}(t+\tau)\right>}{\left<I_{x_\alpha}(t)\right>\left<I_{x_\beta}(t+\tau)\right>}.\\
\end{aligned}
\end{equation}

Here, $\left<\cdots\right>$ denotes time averaging, and $\tau$ represents the time delay between the two photon detection events. From these two correlations, an interferogram $G^{(2)}(\delta_0,\tau)$ can be expressed as
\begin{equation}
\begin{aligned}
G^{(2)}(\delta_0,\tau)=1-\frac{g_{\delta_0,\times}^{(2)}(\tau)}{g_{\delta_0,\parallel}^{(2)}(\tau)}\equiv 1-g^{(2)}_{\delta_0}(\tau).
\end{aligned}
\end{equation}

Finally, the spectral correlation $P(\zeta,\tau)$ is defined as the Fourier transform of the interferogram:
\begin{equation}
\label{Eq.4}
P(\zeta,\tau)={\mathcal{F}}[G^{(2)}(\delta_0,\tau)]_{\delta_0}=\left<\int_{-\infty}^\infty S(\omega,t)\cdot S(\omega+\zeta,t+\tau)d \omega \right>,
\end{equation}
where $S(\omega,t)$ represents the spectral density of the emitter with the time-dependent emission frequency $\omega$ and detection time $t$, $\left<\cdots\right>$ denotes time averaging, and ${\mathcal{F}}[\cdots]_{\delta_0}$ denotes the Fourier transform with respect to $\delta_0$. 
The Fourier transform serves to transform the path-length difference-dependent temporal photon-correlation functions into the frequency-dependent temporal spectral correlation function. A detailed derivation can be found in \cite{brokmann2006photon}. Since the spectral correlation here originates solely from a single emitter, it will be referred as the spectral auto-correlation - not to be mistaken for the intensity auto-correlation $g_{\delta_0,\parallel}^{(2)}(\tau)$.

\subsection{Theory of SFSR imaging——multi-emitters, multi-pixels case}
\label{Sec.2.3}
We continue with the discussion of the spatially resolved image-plane spectral correlations for multiple emitters. The spectral correlations from pixel pairs $x_{\alpha,\beta}$ in the image plane region of high PSF overlap contain terms from single-emitter spectral auto-correlations $P_{||}(\zeta,\tau)$ and multi-emitter cross-correlations $P_{\times}(\zeta,\tau)$. Because the relative contributions to the total spectral correlation scale nonlinearly in the relative emitter intensities, narrower effective PSF can be obtained from the ratio of $P_{||}(\zeta,\tau)$ and $P_{\times}(\zeta,\tau)$. This effective PSF in SFSR imaging is derived as follows.

We consider $N$ emitters that are mapped onto positions in the image plane $x_i$ with $i=1,\cdots,N$. Each emitter appears broadened by a Gaussian PSF centered around $x_i$: 

\begin{equation}
\label{Eq.5}
PSF_i(x)=\frac{1}{2\pi\sigma^2}\exp\left(\frac{-(x-x_i)^2}{2\sigma^2}\right)\equiv h(x-x_i),\quad  \sigma\cong 0.21\cdot \frac{\lambda_i}{NA},
\end{equation} 
where $\lambda_i$ is the emission wavelength and $NA$ is the numerical aperture of the objective lens. At the pixel positions of a detector, $x_\alpha$, with $\alpha=1,\cdots,K$, the observed spectral densities are linear combinations of the spectral densities of all emitters, weighted by their respective PSF at $x_\alpha$:
\begin{equation}
S_{\alpha}(\omega,t)=\mathop\sum_i^N PSF_i(x_{\alpha})  \cdot S_i(\omega,t)\propto\mathop\sum_{i}^Nh(x_{\alpha}-x_i)S_i(\omega,t).
\end{equation}

Hence, the total spectral correlation from two conjugated detectors at two corresponding pixel positions $x_\alpha$ and $x_\beta$ is also a linear combination of weighted spectral auto- and cross-correlations:

\begin{equation}
\begin{aligned}
P_{\alpha\beta}(\zeta,\tau)&=\left<\int_{-\infty}^\infty S_{\alpha}(\omega,t)\cdot S_{\beta}(\omega+\zeta,t+\tau)d \omega \right>\\
&= \mathop\sum_{i}^NPSF_{i}(x_{\alpha})PSF_{i}(x_{\beta})\cdot P_{ii}(\zeta,\tau)\quad &&\cdots\cdots \text{spectral}\: \text{auto-correlation}\\
&+\mathop\sum_{i\neq j}^N PSF_{i}(x_{\alpha})PSF_{j}(x_{\beta})\cdot P_{ij}(\zeta,\tau) &&\cdots\cdots \text{spectral}\: \text{cross-correlation}.
\end{aligned}
\end{equation}

For identical emitters $i$, the spectral auto-correlations $P_{ii}(\zeta,\tau)$ are identical as well: $P_{ii}(\zeta,\tau) \equiv P_{||}(\zeta,\tau)$. Similarly, the spectral cross-correlations $P_{ij}(\zeta,\tau)$, where $i \neq j$, are identical and 
$P_{ij}(\zeta,\tau) \equiv P_{\times}(\zeta,\tau)$. We can then write:

\begin{equation}
\label{Eq.SUM}
\begin{aligned}
P_{\alpha\beta}(\zeta,\tau)= A_{||\alpha\beta}P_{||}(\zeta,\tau) + A_{\times\alpha\beta}P_{\times}(\zeta,\tau),
\end{aligned}
\end{equation}
where we have introduced the amplitudes of the spectral auto- and cross-correlations $A_{||\alpha\beta}$ and $A_{\times\alpha\beta}$, defined as:

\begin{equation}
\label{Eq.8}
\begin{aligned}
A_{||\alpha\beta}&=\mathop\sum_{i}^NPSF_{i}(x_{\alpha})PSF_{i}(x_{\beta})=\mathop\sum_{i}^Nh(x_{\alpha}-x_i)h(x_{\beta}-x_i)\propto \sum_{i=1}^N\exp\left(\frac{-(x_i-\frac{x_{\alpha}+x_{\beta}}{2})^2}{\sigma^2}\right)\\
A_{\times\alpha\beta}&=\mathop\sum_{i\neq j}^N PSF_{i}(x_{\alpha})PSF_{j}(x_{\beta})=\mathop\sum_{i\neq j}^Nh(x_{\alpha}-x_i)h(x_{\beta}-x_j)\\
&\propto \mathop\sum_{i\neq j}^N\exp\left(\frac{-(x_i-x_{\alpha})^2}{2\sigma^2}\right)\exp\left(\frac{-(x_j-x_{\beta})^2}{2\sigma^2}\right)
\end{aligned}
\end{equation}

The second-order SFSR image is the sum over the amplitudes $A_{||\alpha\beta}$ for all $K^2$ pixel pairs $\alpha, \beta$ between the two detector arrays. The corresponding effective PSF, $U^{(2)}_{SFSR}$, is then given by the following expression:

\begin{equation}
\label{Eq.9}
U^{(2)}_{SFSR}\propto \mathop\sum_{\alpha,\beta=1}^K\mathop\sum_{i=1}^N\exp\left(\frac{-(x_i-\frac{x_{\alpha}+x_{\beta}}{2})^2}{\sigma^2}\right).
\end{equation}

This expression is similar in form to those obtained for SOFI and AM \cite{dertinger2009fast,PhysRevA.85.033812,schwartz2013super}, indicating the same level of resolution enhancement by a factor of $\sqrt2$ for second-order SFSR and ICM. Like ICM approaches, SFSR can be extended to the $n^{th}$ order, theoretically providing a $\sqrt{n}$-fold resolution enhancement compared to WFM. The complete derivation is shown in section \ref{SI.S1}, Supplement 1. For practical SFSR image reconstruction, both amplitudes $A_{||}$ and $A_{\times}$ can be extracted by fitting the total spectral correlation according to Eq. \ref{Eq.SUM}.

To conclude our derivation, we discuss the special case of exactly $N=2$ emitters imaged with second-order SFSR. In this case, an additional $\sqrt 2$-fold resolution enhancement can be gained by using the cross-term elimination (XE) method. For $N=2$, Eq. \ref{Eq.8} provide the following expressions for $A_{||}$ and $A_{\times}$:

\begin{equation}
\begin{aligned}
&{{A_{||}}}&&=PSF_1^2+ PSF_2^2\propto \exp\left(\frac{-(x_1-\frac{x_{\alpha}+x_{\beta}}{2})^2}{\sigma^2}\right)+\exp\left(\frac{-(x_2-\frac{x_{\alpha}+x_{\beta}}{2})^2}{\sigma^2}\right)\\
&{{A_{\times}}}&&=2PSF_1 PSF_2\propto 2\exp\left(\frac{-(x_1-\frac{x_{\alpha}+x_{\beta}}{2})^2}{\sigma^2}\right)\exp\left(\frac{-(x_2-\frac{x_{\alpha}+x_{\beta}}{2})^2}{\sigma^2}\right).
\end{aligned}
\end{equation}

SFSR-XE images can be similarly reconstructed after pixel-reassignment and summation over all $K^2$ intensity images, but this time by taking $A_{||}^2-\frac{1}{2}A_{\times}^2$ as the contrast function. This XE approach eliminates the non-vanishing spectral cross-correlation contributions. The effective PSFs for SFSR-XE reconstructed images, $U^{(2)}_{SFSR-XE}$, can be expressed as:
\begin{equation}
\label{Eq.11}
\begin{aligned}
A_{||}^2-\frac{1}{2}A_{\times}^2&=PSF_1^4+ PSF_2^4=(PSF_1^2+ PSF_2^2)^2-2PSF_1^2 PSF_2^2\\
&\propto \exp\left(\frac{-2(x_1-\frac{x_{\alpha}+x_{\beta}}{2})^2}{\sigma^2}\right)+\exp\left(\frac{-2(x_2-\frac{x_{\alpha}+x_{\beta}}{2})^2}{\sigma^2}\right),\\
U^{(2)}_{SFSR-XE}&\propto \mathop\sum_{\alpha,\beta=1}^K\mathop\sum_{i=1}^2\exp\left(\frac{-2(x_i-\frac{x_{\alpha}+x_{\beta}}{2})^2}{\sigma^2}\right),\\
\end{aligned}
\end{equation}
which shows the additional $\sqrt2$-fold resolution enhancement. The total resolution advantage over WFM is therefore a factor of $2$,  but only for strictly two independent emitters.

\section{Numerical Simulations of SFSR imaging}
\label{sec.3}
To demonstrate the SFSR image reconstruction and show its expected signal-to-noise ratio, we conducted numerical statistical optics simulations for realistic emitter brightness, adapted from \cite{utzat2021lifetime} (see section S3, Supplement 1). A generic model for single emitter spectral fluctuations was chosen, which describes stochastic discrete frequency hopping of a narrow homogeneous Lorentzian lineshape $S(\omega+\omega_{0}(t))$ to different center wavelengths $\omega_{0}(t)$, where the bounds of $\omega_{0}(t)$ are defined by a Gaussian envelop \cite{utzat2021lifetime}. The fully diffused lineshape is therefore the convolution of the Gaussian and the homogeneous Lorentzian lineshape. The model was parametrized in the characteristic spectral jumping time $\tau_c$, the homogeneous Lorentzian linewidth, and the width of the Gaussian probability distribution. Concrete simulation parameters were chosen from a range of experimental values for quantum emitters in hexagonal boron nitride (hBN) and quantum dots (purely homogeneous linewidths of an emitter in hBN, a CdSe quantum dot, and a nitrogen-vacancy center are 5 $\mu eV$, 50 $meV$, and 0.05 $\mu eV$, respectively, while their time-averaged, spectrally diffused linewidths are 50 $\mu eV$, 100 $meV$, and 0.5 $meV$ \cite{chen2013compact,spokoyny2020effect,yurgens2021low}).

\begin{figure}[ht]
\centering\includegraphics[width=8cm]{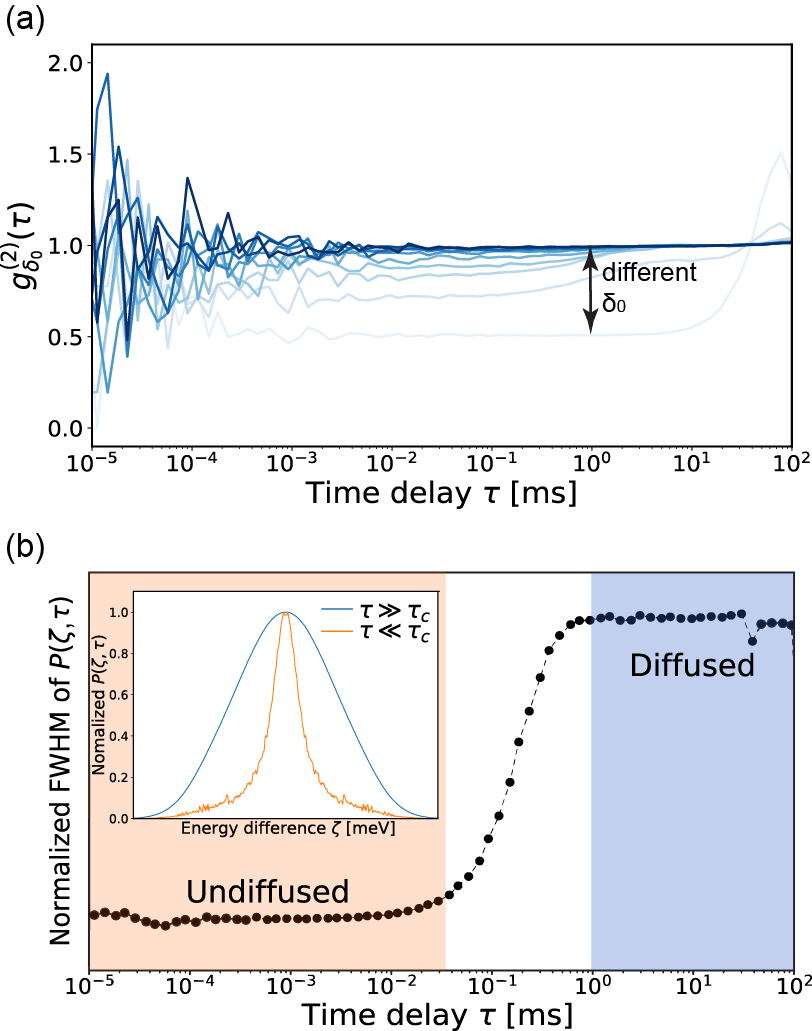}
\caption{Statistical optics Monte Carlo simulation of PCFS. (a) $g_{\delta_0}^{(2)}(\tau)$ for different $\delta_0$ indicating an infletion at the characteristic spectral jumping time. (b) Evolution of the FWHM of $P(\zeta,\tau)$ with $\tau$. The width increases from $\tau\ll\tau_c$ (undiffused) to $\tau\gg\tau_c$ (diffused). Inset: Diffused and undiffused $P(\zeta,\tau)$ obtained via the Fourier transform of $g_{\delta_0}^{(2)}(\tau)$.}
\label{Fig.3}
\end{figure}

The simulation results for the single-emitter and single-pixel case, i.e., PCFS, are shown in Fig. \ref{Fig.3}. The observable in PCFS has been discussed in detail elsewhere \cite{brokmann2006photon,PhysRevLett.100.027403,Coolen:09,utzat2021lifetime}. From the perspective of SFRS imaging, we point out the evolution of $g^{(2)}_{\delta_0}(\tau)$. Akin to SOFI, an inflection at $\tau_{c}$ indicates the characteristic fluctuation time, here the average spectral jumping time. The same inflection is observed in the plot of the full-width at half maximum (FWHM) of the single-emitter spectral correlation $P(\zeta,\tau)$ in Fig. \ref{Fig.3}b, which was obtained according to Eq. \ref{Eq.4}. We note that the spectral correlation is equivalent to the spectral auto-correlation $P_{||}(\zeta,\tau)$ in the case of a single emitter here. The observed dynamic broadening is attributed to the loss of spectral correlation of emission from the same emitter at the characteristic spectral jumping time. The corresponding spectral correlations for $\tau < \tau_c$ and $\tau > \tau_c$ (inset) reflect the undiffused Lorentzian and diffused Gaussian, respectively.

\begin{figure}[h]
\centering\includegraphics[width=10cm]{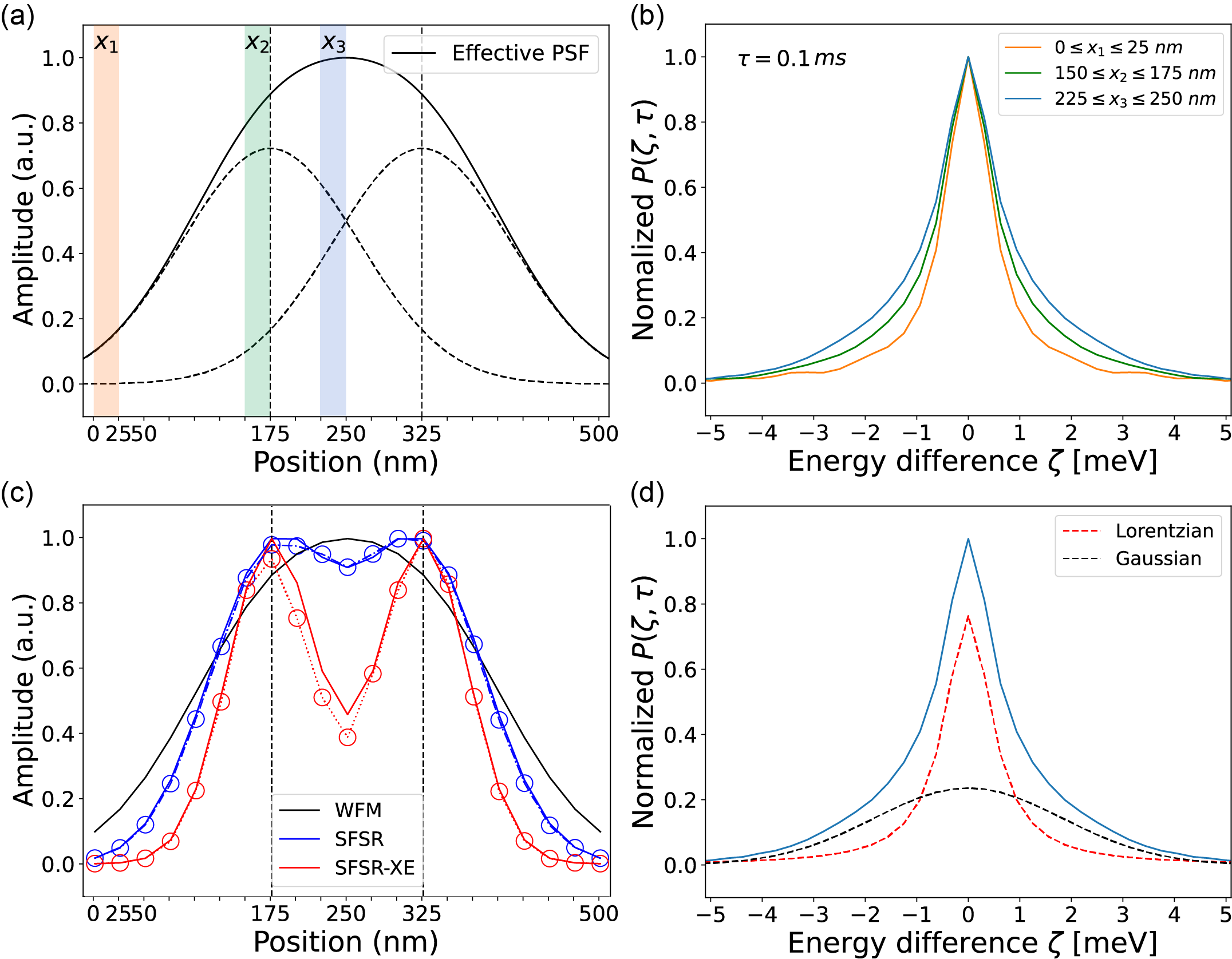}
\caption{Simulated effective PSF for a 1D SFSR image. (a) Overlapping PSFs of diffraction-limited WFM images of two closely situated emitters. The vertical dotted lines at $x_1=175 \ nm$ and $x_2=325 \ nm$ indicate the position of the two emitters. (b) Three selected $P_{\alpha,\beta}(\zeta,\tau)$ at different pixel positions indicated by the three shaded areas in (a). (c) Reconstructed PSFs of SFSR from fitted $A_{\parallel}$ and $A_{\times}$. The fitting process is shown in (d). The simulated effective PSFs of SFSR and SFSR-XE are shown by the dotted curves and compared to the theory shown by the solid curves. (d) The summed spectral correlation pattern at $x=250 \ nm$ is shown by the blue dotted curve. Red and black dotted curves show the fitting result of the Lorentzian and Gaussian components, respectively.}
\label{Fig.4}
\end{figure}

\vspace{5pt}
Extended simulations for the case of two emitters exhibiting independent spectral fluctuations demonstrate the SFSR resolution advantage. The simulation in Fig. \ref{Fig.4} mimics two emitters located at $x_1=175 \ nm$ and $x_2=325 \ nm$ in the object plane. For simplicity, we super-impose the image plane implying a magnification of one. The intensity profile of each emitter follows a Gaussian PSF according to Eq. \ref{Eq.5} with $NA=1.2$ and a total of $10^6$ photons per second are detected per emitter.  Each emitter has a center wavelength of 500 $nm$, Lorentzian homogeneous linewidth of 0.5 $meV$, a Gaussian width of 2.35 $meV$ as a bound for spectral jumping, and a time-invariant probability of spectral jumping with an average jump time of $\tau_{c}=1$ $ms$.

The 1D image is recorded with 20 hypothetical pixels, correspondingly collecting from 25 $nm$ increments in the object plane. The number of photons received per emitter by a given pixel scales with the value of the respective Gaussian PSFs. We simulate the pairwise spectral correlations for conjugated detector pairs with a time of integration of 1 $s$, with the same relative positions $x_\alpha=x_\beta$ in the image plane, excluding all $x_\alpha \neq x_\beta$. This reduces the computational complexity from  $K^2$ to $K$ according to Eq. \ref{Eq.9}. This step is purely methodological and does not affect the final SFSR resolution. In real imaging scenarios, computation of all pairwise spectral correlations will increase the SNR of the contrast function for a given total photon budget.

\begin{figure}[ht]
\centering\includegraphics[width=10cm]{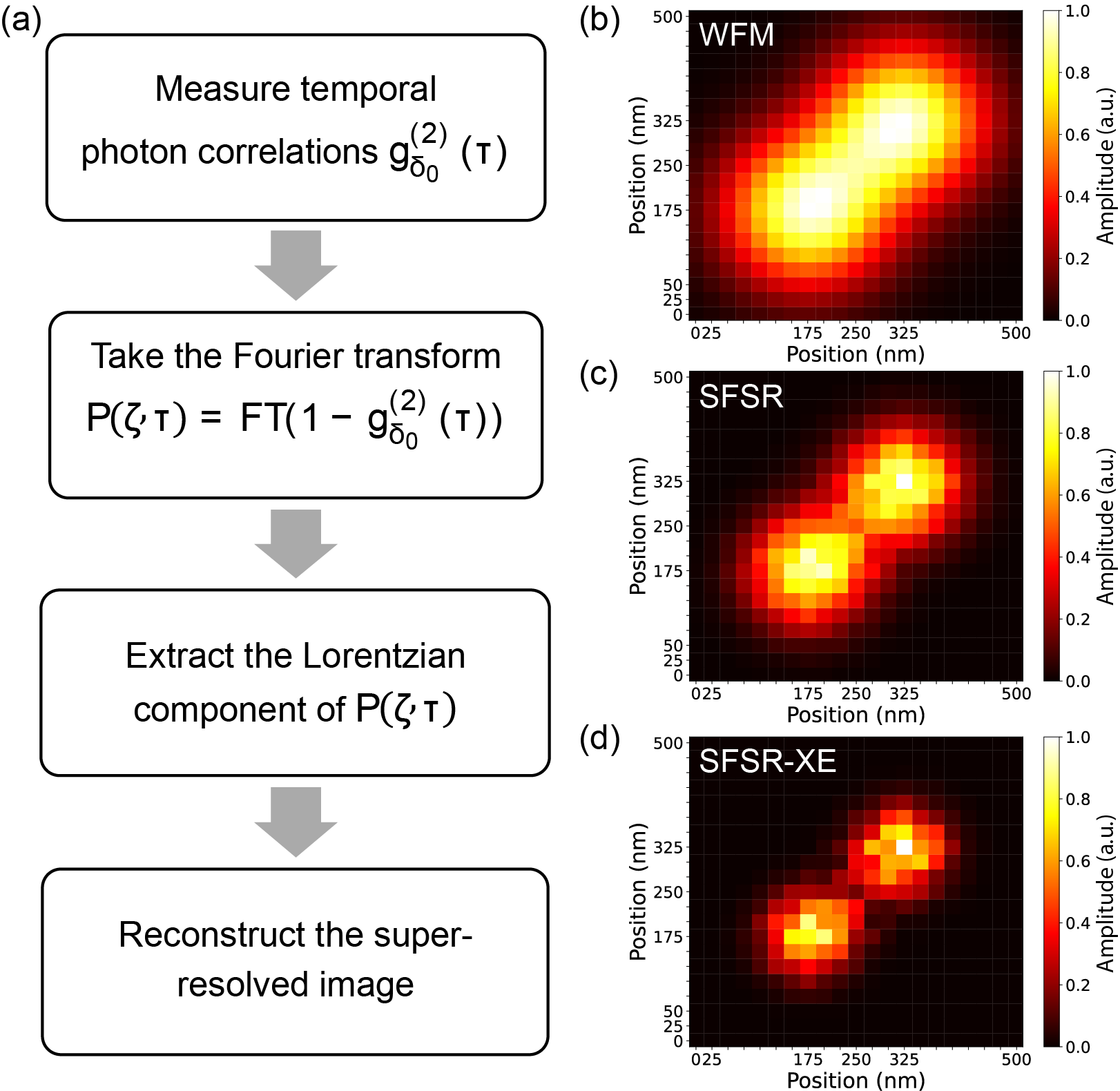}
\caption{2D visualization of resolution enhancement. (a) Flow chart of SFSR image reconstruction. (b) 2D WFM image of two emitters located at $x_1 = 175 \ nm, y_1=175 \ nm$ and $x_2=325 \ nm, y_2=325 \ nm$. (c) 2D SFSR image of the same emitter configuration as for (b). (d) 2D SFSR-XE image of the same emitter system as for (b).}
\label{Fig.5}
\end{figure}

SFSR's contrast function is based on the modulation of the single-emitter spectral auto-correlation amplitudes $A_{||}$ across the image plane. Because $A_{||}$ decays with the characteristic fluctuation time, $P_{\alpha,\beta}(\zeta,\tau)$ for $\tau<\tau_{c}$ must be considered for image reconstruction. Fig. \ref{Fig.4}b shows three corresponding $P_{\alpha,\beta}(\zeta,\tau)$ collected from different pixel positions $x_{\alpha,\beta}$, with $\alpha=\beta \in \{1-20\}$. The total $P_{\alpha,\beta}(\zeta,\tau)$ are linear combinations of the spectral auto- and cross-correlations (see Eq. \ref{Eq.SUM}), and therefore exhibit different shapes. Pixel-wise quantification of $A_{||}$ is achieved via a fit of Lorentzian and Gaussian spectral correlations, as shown for one pixel in Fig. \ref{Fig.4}d. SFSR and SFSR-XE images are then reconstructed from $A_{||}$ and $A_{\times}$ according to Eq. \ref{Eq.9} and \ref{Eq.11}, respectively. The effective PSFs of both SFSR and SFSR-XE are compared with that of the WFM base case in Fig. \ref{Fig.4}c. The narrowed effective PSFs manifest as distinct effective Gaussian features, thus demonstrating the improvement in resolution compared to the WFM image. The observed effective PSF narrowing is in line with our theoretical prediction, confirming the expected $\sqrt{2}$ and $2$-fold enhancements for SFSR and SFSR-XE imaging, respectively. The same resolution advantage can be seen in simulated 2D SFSR and SFSR-XE images with the same spectral fluctuation parameters in Fig. \ref{Fig.5}b-d.

\vspace{5pt}

\begin{figure}[ht]
\centering\includegraphics[width=12cm]{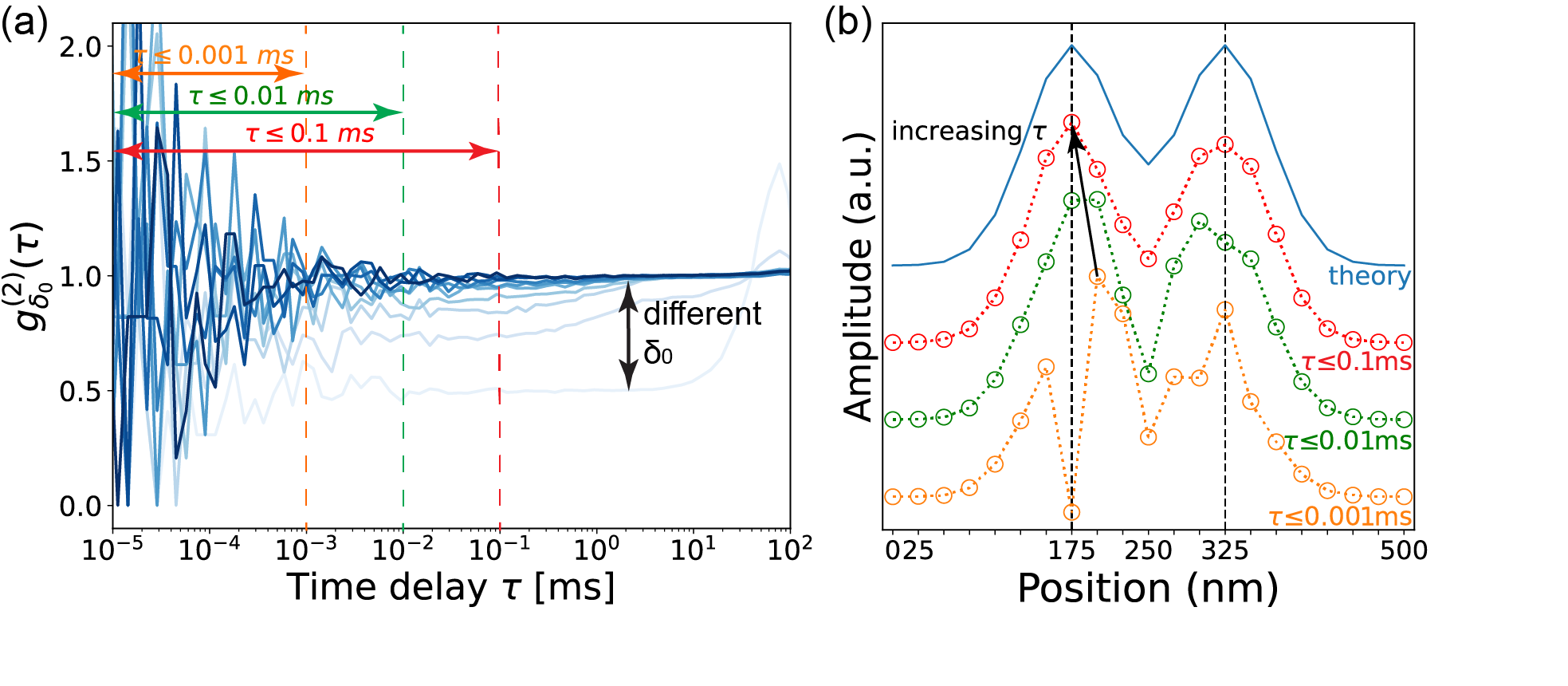}
\caption{Improved SNR of effective PSFs of SFSR-XE through temporal averaging. (a) Illustration of the different $\tau$-intervals in $g_{\delta_0}^{(2)}(\tau)$ at different $\delta_0$ used for the SFSR-XE image reconstruction. (b) Reconstructed SFSR-XE effective PSFs with $\tau$-interval of 0.001, 0.01, and 0.1 $ms$ indicating improvements in image quality. Effective PSFs are vertically offset for better visual illustration. The dotted lines at $x_1=175 \ nm$ and $x_2=325 \ nm$ indicate the position of the two emitters.}
\label{Fig.6}
\end{figure}

Finally, we demonstrate a method to maximize the SNR in SFSR imaging. SFSR imaging can in principle utilize all photon pairs with $\tau < \tau_c$ to distinguish between the undiffused spectral auto-correlations and the multi-emitter cross-correlations. To illustrate this advantage, we repeated the simulation of 2D SFSR images with lower photon count rates of $10^5$ photons per second per emitter. SFSR-XE images were reconstructed using all photons pairs with $\tau \leq 0.001ms$, $\tau \leq 0.01ms$, and $\tau \leq 0.1ms$ from the intensity correlations in Fig. \ref{Fig.6}a. These are as shown in Fig. \ref{Fig.6}b. The improved image quality with longer $\tau$-intervals can be both visually assessed and quantified by the reduced deviation of the reconstructed effective PSF maxima from actual emitter positions. This deviation towards the overlapping region of the emitter PSFs occurs under conditions of low photon counts, short time of integration, and short $\tau$-interval due to the elevated noise, as detailed in section S4, Supplement 1.

\section{Discussion}
The above derivation and simulations suggest SFSR as a well-suited method for imaging of identical emitters, such as quantum emitters at low temperatures. We discuss the use of the spectral correlation function in SFSR within the context of existing methods.

The inclusion of spectral information in super-resolution microscopy per se is not new. STORM and PALM were successfully combined with a prism or spectrometer to correlate both spatial and spectral channels, which enabled spectral separation of multiple fluorescent labels and hence achieved spectrally resolved super-resolution microscopy \cite{zhang2015ultrahigh,mlodzianoski2016super}. For molecular sub-ensembles with inhomogeneous broadening much larger than the homogeneous linewidth, photoluminescence excitation can select individual molecules, first proposed by Betzig \cite{betzig1995proposed}. Experimental demonstration of this idea were conducted by Oijen \textit {et al.}, by tuning a narrow band laser into resonance with the absorbance line of a single emitter \cite{van19983}; and by Boschetti \textit {et al.}, using a random laser featuring sharp spikes at uncorrelated frequencies to sparsely sample the inhomogeneous linewidth \cite{boschetti2020spectral}. However, this approach cannot be used in cases where the spectral diffusion amplitude is larger than the inhomogeneous broadening and when the spectral fluctuation happens on timescales faster than the spectral acquisition time. None of these methods utilizes spectral correlations as the image contrast.

SFSR is conceptually different. It does not primarily use spectral information to extract further sample information, although this is possible as discussed below. It also does not rely on spectral sorting and stochastic reconstruction, which would in essence be a spectral analog to STORM. Rather, SFSR uses spectral correlation functions as an image reconstruction contrast, which is directly analogous to intensity correlation functions in SOFI. Similar to SOFI and AM, our SFSR technique can be extended to higher-order correlations, in theory achieving a $\sqrt{n}$-fold resolution improvement. Furthermore, SFSR imaging can be integrated with image scanning microscopy (ISM) \cite{muller2010image} as shown in section S2, Supplement 1, akin to SOFISM \cite{sroda2020sofism} and Q-ISM \cite{tenne2019super}, which combine SOFI and AM with ISM, respectively. This ISM integration theoretically provides a $\sqrt{2n}$-fold improvement in resolution over WFM. Incorporating ISM involves scanning the object plane and reconstructing the final image through pixel reassignment, which aligns well with our proposed experimental setup. Additionally, the application of Fourier reweighting and machine learning techniques in post-processing can be explored to further enhance image quality. Fourier reweighting filters out the oscillatory higher frequency Fourier modes of the image for better image contrast \cite{muller2010image}.  Machine learning can optimize the fitting of spectral correlations and improve the accuracy of emitter localization, thereby possibly boosting the performance of the SFSR technique \cite{beck2024improving,proppe2023time}.

The SFSR approach also enables simultaneous SOFI and AM images reconstruction without any additional data acquisition or longer integration time. For this the SOFI and (or) AM correlation analysis can be performed on the sum signal of both conjugate ports of the interferometer, in which case the interferometer will not significantly affect the analysis as long as the path-length difference is much shorter than the characteristic times of intensity fluctuation or anti-bunching. SOFI, AM, and SFSR images can then be generated using the up to $n^{th}$ order temporal photon correlation.

This inherent multi-modality of SFSR super-resolution uses the quantum emission statistics, intensity fluctuations, and spectral fluctuations and may be advantageous in several ways. First, multi-dimensional contrast functions will improve the accuracy of emitter localization. Second, orthogonality in contrast function variables may help to study new nanoscale physics. For example, the independent localization of two emitters using antibunching microscopy and SFSR may reveal the degree to which two separated emitters couple to the same or different bath fluctuations causing spectral diffusion, thereby quantifying the spatial extent of emitter-bath coupling.

Finally, SFSR is not bound to any specific stochastic form of emitter spectral fluctuations and can work with a range of emitters undergoing e.g., continuous spectral random-walking, exemplified by the Wiener model \cite{mazo2008brownian,nitzan2024chemical}, or discrete spectral jumping \cite{beyler2013direct}. As long as the emitters show independent spectral fluctuations, and the spectral auto- and cross-correlations can be distinguished, SFSR can achieve super-resolution. For most quantum emitters at low temperatures, the undiffused single-emitter spectral correlation can be much narrower than the diffused lineshape, sometimes by two orders of magnitude \cite{beyler2013direct,yurgens2021low,adl2021homogeneous}, which will support the reconstruction of images.

\section{Conclusions}
In conclusion, we theoretically propose SFSR as a new super-resolution concept that provides a $\sqrt{2}$-fold resolution advantage for emitters undergoing independent spectral fluctuations on arbitrary timescales. A 2-fold resolution enhancement can be achieved for scenes with exactly two emitters. Numerical simulations confirm our theoretical results and demonstrate the practicality of resolving quantum emitters with literature-informed spectral diffusion amplitudes and characteristic jumping timescales. Our concept may therefore be useful in imaging of high-quality quantum emitters that are designed to be spectrally indistinguishable but for which spectral fluctuations are notoriously difficult to eliminate. SFSR may apply to use cases complementary to SOFI. At room temperature, emitters often exhibit intensity fluctuations and broad homogeneous linewidths with comparatively little spectral diffusion, favoring the use of SOFI over SFSR. At low temperature, intensity intermittency is often suppressed while spectral diffusion persists, often over multiples of the homogenous linewidth, suggesting SFSR as an alternative.

\section*{Funding}
This work was supported by the Laboratory Directed Research and Development Program of Lawrence Berkeley National Laboratory under U.S. Department of Energy Contract No. DE-AC02-05CH11231. We gratefully acknowledge the generous support from UC Berkeley and the UC Berkeley College of Chemistry.

\section*{Acknowledgments}
This research used the Savio computational cluster resource provided by the Berkeley Research Computing program at the University of California, Berkeley (supported by the UC Berkeley Chancellor, Vice Chancellor for Research, and Chief Information Officer).

\section*{Disclosures}
The authors declare no conflict of interest.
\section*{Data availability}
Data underlying the results presented in this paper may be obtained from the authors upon reasonable request.
\section*{Supplemental document}
See Supplement 1 for supporting content.

\newpage
\section*{Supplementary materials}

\renewcommand{\thesection}{S\arabic{section}}
\renewcommand{\thesubsection}{S\arabic{subsection}}
\renewcommand{\theequation}{S\arabic{equation}}
\renewcommand{\thetable}{S\arabic{table}}
\renewcommand{\thefigure}{S\arabic{figure}}

\setcounter{section}{1} 
\setcounter{equation}{0}
\setcounter{table}{0}
\setcounter{figure}{0}

\subsection{Theory of higher order SFSR imaging}
\label{SI.S1}
SFSR imaging can be extended to the $n^{th}$ order by considering the resulting spectral correlation from $n$ conjugated detectors at $n$ pixel positions, $x_\alpha,x_\beta,\cdots,x_n$:
\begin{equation}
\label{Eq.S1}
\begin{aligned}
P_{\alpha\beta\cdots n}(\zeta,\tau)&=\left<\int_{-\infty}^\infty S_{\alpha}(\omega,t)\cdot S_{\beta}(\omega+\zeta_\beta,t+\tau_\beta)\cdots S_{n}(\omega+\zeta_n,t+\tau_n) d \omega \right>\\
&= \mathop\sum_{i_1=i_2=\cdots=i_n}^NPSF_{i_1}(x_{\alpha})PSF_{i_2}(x_{\beta})\cdots PSF_{i_n}(x_{n}) \cdot P_{i_1i_2\cdots i_n}(\zeta,\tau)\\
&+\mathop\sum_{any\: i_j\neq i_k}^N PSF_{i_1}(x_{\alpha})PSF_{i_2}(x_{\beta})\cdots PSF_{i_n}(x_{n}) \cdot P_{i_1i_2\cdots i_n}(\zeta,\tau).
\end{aligned}
\end{equation}

As before, the first summation term represents the spectral auto-correlation contribution, and the second summation term represents the spectral cross-correlation contribution. Only the spectral auto-correlation term ($i_1=i_2=\cdots=i_n$) from the emission of the same emitter gives a high-dimensional spectral auto-correlation. Any spectral cross-correlation term ($i_j\neq i_k$) will give a broader shape, due to the lack of correlation between two independently diffusing emitters.  Hence, the separation of these two  component from the total spectral correlation is still possible by fitting the total according to Eq. \ref{Eq.S1}. The acquired amplitude of the spectral auto-correlation can be expressed as:

\begin{equation}
\begin{aligned}
A_\parallel&=\mathop\sum_{i_1=i_2=\cdots=i_n}^NPSF_{i_1}(x_{\alpha})PSF_{i_2}(x_{\beta})\cdots PSF_{i_n}(x_{n}) \\
&=\mathop\sum_{i}^Nh(x_{\alpha}-x_i)h(x_{\beta}-x_i)\cdots h(x_{n}-x_i)\\
&\propto \sum_{i=1}^N\exp\left(\frac{-n(x_i-\frac{x_{\alpha}+x_{\beta}+\cdots+x_n}{n})^2}{2\sigma^2}\right).
\end{aligned}
\end{equation}

Similarly, the final pixel-reassigned images can be reconstructed after pixel-reassignment and summation over all images collected from any combination of $n$ detectors. Its effective PSF ($U^{(n)}_{SFSR}$) can be expressed as:
\begin{equation}
U^{(n)}_{SFSR}\propto \mathop\sum_{\alpha,\beta,\cdots,n=1}^K\mathop\sum_{i=1}^N\exp\left(\frac{-n(x_i-\frac{x_{\alpha}+x_{\beta}+\cdots+x_n}{n})^2}{2\sigma^2}\right),
\end{equation}
which gives a theoretical $\sqrt{n}$-fold resolution enhancement compared to WFM.

\subsection{Theory of SFSR-ISM}

To showcase the compatibility of SFSR imaging with image scanning microscopy (ISM), the following derivation is conducted under the ISM setup by adding a scanning stage into the original SFSR imaging setup. This offers an additional $\sqrt 2$-fold resolution enhancement \cite{muller2010image}. In addition to the previous assumptions of PCFS (see section \ref{Sec.2.3}), we assume both laser excitation and emitter emissions have Gaussian profiles:
\begin{equation}
PSF_i(x)=\frac{1}{2\pi\sigma^2}\exp\left(\frac{-(x-x_i)^2}{2\sigma^2}\right)=h(x-x_i),\quad  \sigma\cong 0.21\cdot \frac{\lambda_i}{NA},
\end{equation} 
where $\lambda_i$ is the wavelength and $NA$ is the numerical aperture of the objective.

The PSFs of $N$ emitters mapped onto  positions $x_i, i=1,\cdots,N$ on the image plane, now involve the PSFs of the excitation:

\begin{equation}
PSF_i(x)=h_{exc}(x_s-x_i)h(x-(x_s-x_i)).
\end{equation}

At pixel position $x_\alpha, \alpha=1,\cdots,K$, the effective spectral densities is a linear combination of the spectral densities of all emitters, weighted by their PSFs at $x_\alpha$:
\begin{equation}
S_{\alpha}(\omega,t)=\mathop\sum_i^N PSF_i(x_{\alpha})  \cdot S_i(\omega,t)\propto\mathop\sum_{i}^Nh_{exc}(x_s-x_i)h(x_{\alpha}-(x_s-x_i))S_i(\omega,t).
\end{equation}

The resulting spectral correlation from two detectors corresponding to the pixel position $x_\alpha$ and $x_\beta$ again can be written as:

\begin{equation}
\begin{aligned}
P_{\alpha\beta}(\zeta,\tau)&=\left<\int_{-\infty}^\infty S_{\alpha}(\omega,t)\cdot S_{\beta}(\omega+\zeta,t+\tau)d \omega \right>\\
&= \mathop\sum_{i}^NPSF_{i}(x_{\alpha})PSF_{i}(x_{\beta})\cdot P_{ii}(\zeta,\tau)\quad &&\cdots\cdots \text{spectral}\: \text{auto-correlation}\\
&+\mathop\sum_{i\neq j}^N PSF_{i}(x_{\alpha})PSF_{j}(x_{\beta})\cdot P_{ij}(\zeta,\tau) &&\cdots\cdots \text{spectral}\: \text{cross-correlation}.
\end{aligned}
\end{equation}

The contributions of spectral auto- and cross-correlations can be similarly separated  with the amplitudes of the spectral auto- and cross correlations ($A_\parallel,A_\times$) to the total spectral correlation:
\begin{equation}
\begin{aligned}
A_\parallel&=\mathop\sum_{i}^NPSF_{i}(x_{\alpha})PSF_{i}(x_{\beta})=\mathop\sum_{i}^Nh_{exc}^2(x_s-x_i)h(x_{\alpha}-(x_s-x_i))h(x_{\beta}-(x_s-x_i))\\
& \propto \sum_{i=1}^N\exp\left(\frac{-2(x_s-(x_i-\frac{x_{\alpha}+x_{\beta}}{4}))^2}{\sigma^2}\right)\\
A_\times&=\mathop\sum_{i\neq j}^N PSF_{i}(x_{\alpha})PSF_{j}(x_{\beta})\\
&=\mathop\sum_{i\neq j}^Nh_{exc}(x_s-x_i)h(x_{\alpha}-(x_s-x_i))h_{exc}(x_s-x_j)h(x_{\beta}-(x_s-x_j))\\
&\propto \mathop\sum_{i\neq j}^N\exp\left(\frac{-(x_s-(x_i-\frac{1}{2}x_{\alpha}))^2}{\sigma^2}\right)\exp\left(\frac{-(x_s-(x_j-\frac{1}{2}x_{\beta}))^2}{\sigma^2}\right).
\end{aligned}
\end{equation}

The second-order SFSR image is then reconstructed by assigning the amplitude of the Lorentzian as the intensity at pixel-reassigned position $\frac{x_\alpha+x_\beta}{4}$, and summing all $K^2$ intensity images created by correlating every pairs of detectors. Overall, the final reconstructed image has an effective PSF of is again the exact same form as the SOFISM and Q-ISM \cite{sroda2020sofism,tenne2019super}:
\begin{equation}
U^{(2)}_{SFSR}\propto \mathop\sum_{\alpha,\beta=1}^K\mathop\sum_{i}^N\exp\left(\frac{-2(x_s-(x_i-\frac{x_{\alpha}+x_{\beta}}{4}))^2}{\sigma^2}\right).
\end{equation}

The approach can also be extended to the $n^{th}$ order, which gives a theoretical $\sqrt{2n}$-fold resolution enhancement compared to WFM.

Under ISM setup, the cross-term elimination still offers an additional $\sqrt2$-fold resolution enhancement when imaging a two-emitter system:

\begin{equation}
\begin{aligned}
&{{A_\parallel}}&&=PSF_1^2+ PSF_2^2\\
&&&\propto \exp\left(\frac{-2(x_s-(x_1-\frac{1}{2}x_{\alpha}))^2}{\sigma^2}\right)+\exp\left(\frac{-2(x_s-(x_2-\frac{1}{2}x_{\alpha}))^2}{\sigma^2}\right)\\
&{{A_\times}}&&=2PSF_1 PSF_2\\
&&&\propto 2\exp\left(\frac{-(x_s-(x_1-\frac{1}{2}x_{\alpha}))^2}{\sigma^2}\right)\exp\left(\frac{-(x_s-(x_2-\frac{1}{2}x_{\alpha}))^2}{\sigma^2}\right)\\
&U^{(2)}_{SFSR}&&=PSF_1^4+ PSF_2^4=(PSF_1^2+ PSF_2^2)^2-2PSF_1^2 PSF_2^2={A_\parallel}^2-\frac{1}{2}{A_\times}^2\\
&&&\propto \exp\left(\frac{-4(x_s-(x_1-\frac{1}{2}x_{\alpha}))^2}{\sigma^2}\right)+\exp\left(\frac{-4(x_s-(x_2-\frac{1}{2}x_{\alpha}))^2}{\sigma^2}\right).
\end{aligned}
\end{equation}

Overall, SFSR-ISM-XE can give up to a $2\sqrt{2}$-fold resolution enhancement compared to WFM.

\subsection{Method of PCFS and SFSR simulations}

Given the setup of the simulation discussed in section \ref{sec.3}, the second-order temporal and spectral correlations of SFSR can be generated as follows. At each pixel, we generated a stream of emitted photons from two emitters, with photon-statistics following a Poisson distribution and frequency distribution determined by the generic model that describes emitter spectral fluctuation (see section \ref{sec.3}). The number of photons generated from each emitter equal to the photon counts per second per emitter weighted by the amplitude to their PSFs at that pixel. Log-space temporal photon correlations, $g^{(2)}_{\delta_0,\parallel}(\tau)$ and $g^{(2)}_{\delta_0,\times}(\tau)$, are computed under varying $\delta_0$ with time of integration of 1 $s$. $P(\zeta,\tau)$ on the undiffused timescale ($\tau = 0.1 \ ms$), are then obtained through the fast Fourier transform, as described by Eq. \ref{Eq.4}.  PCFS, as the single-emitter, single-pixel version of SFSR, can be simulated in the same manner, by generating photons from only one emitter and computing correlations at one specific pixel.

The fitting of the total spectral correlation with the sum of a narrow Lorentzian and a broader Gaussian is conducted via the Python SciPy package \cite{2020SciPy-NMeth}. The fitting variables includes the amplitudes, the widths, and the center positions of the Lorentzian and Gaussian peaks. The amplitudes of the fitted Lorentzian and Gaussian peaks at each pixel are recorded as $A_\parallel$ and $A_\times$ respectively. $A_\parallel$ and $A_\times$ are normalized based on the photon counts at these pixels. $A_\parallel$ are assigned as the amplitude at each pixel to form the reconstructed SFSR images, while ${A_\parallel}^2-\frac{1}{2}{A_\times}^2$ are assigned as the amplitude of the SFSR-XE images. In the case where Lorentzian and Gaussian linewidths are close, the Gaussian spectral correlation used in fitting is replaced by a Voigt profile that accurately reflects its shape.

We also simulated the scenario where the two emitters have different emission center wavelengths, linewidths, and characteristic spectral jumping times under the framework of the mentioned generic model. By keeping the positions of the two emitters fixed, no visible change in the reconstructed SFSR image was observed. Table \ref{table.S1} summarizes the tunable parameters in the simulation.

\begin{table}[ht]
\caption{Tunable parameters of numerical simulation of SFSR imaging}
\label{table.S1}
\centering
\begin{tabular}{@{}cc@{}}
\toprule
Tunable parameters       & Our simulation                      \\ 
\midrule
Numerical aperture       & 1.2              \\
Time of integration      & 1 $s$               \\
Emitter position         & $x_1=175 \ nm$, $x_2=325 \ nm^\ast$\\
Photon count rate & $10^6$ $cps$                \\
Emission center wavelength        & 500 $nm$                \\
Lorentzian homogeneous linewidth     & 0.5 $meV$         \\
Gaussian inhomogeneous linewidth   & 2.35 $meV$          \\
Characteristic spectral jumping time  & 1 $ms$         \\ 
\bottomrule
\multicolumn{2}{p{10cm}}{\footnotesize $\ast$ For the 2D simulation, the two emitters are positioned at $x_1 = 175 \ nm, y_1=175 \ nm$ and $x_2=325 \ nm, y_2=325 \ nm$.}\\
\end{tabular}
\end{table}

The exact position of emitters are localized by fitting two Gaussians again via the Python SciPy package \cite{2020SciPy-NMeth}.  The fitting variables include the widths and the center positions of the Gaussians. Center positions of both Gaussians and the fitting errors are recorded, as a measure of the performance of the localization.

\subsection{Systematic error of emitter localization at low photon count rate and short time of integration}

\begin{figure}[ht]
\centering\includegraphics[width=13.3cm]{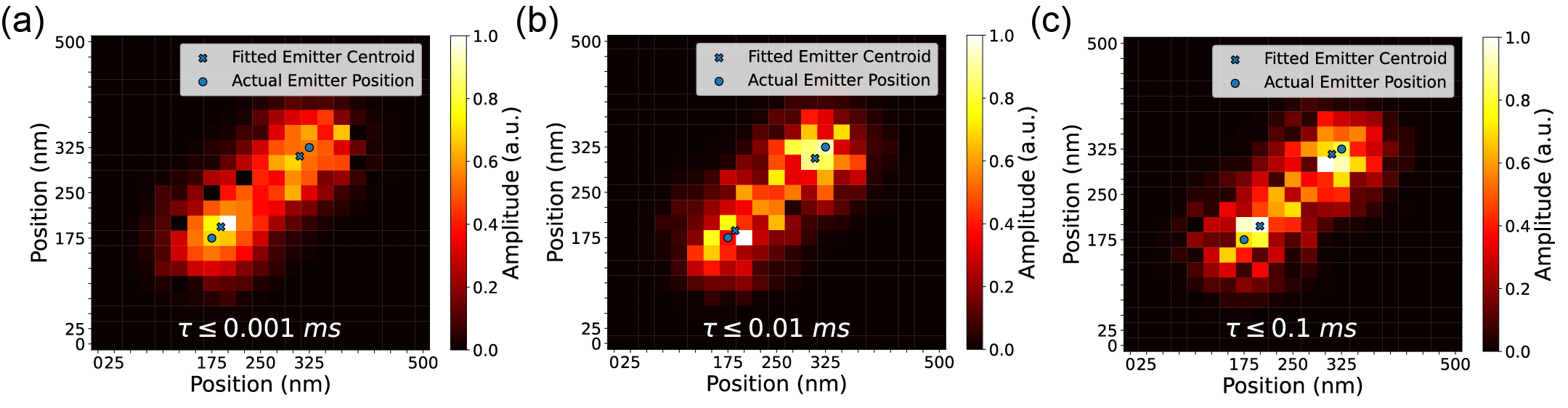}
\caption{Reconstructed SFSR-XE images with $\tau$-interval of 0.001, 0.01, and 0.1 $ms$ for the undiffused spectral correlation. (a-c) Systematic error of emitter localization. The fitted emitter centroids and actual emitter positions are shown by the blue circle and cross, respectively. Consistent deviation towards the overlapping region of the emitter PSFs is observed.}
\label{Fig.7}
\end{figure}

Systematic errors of emitter localization, exceeding the fitting error, arise when the photon count rate is reduced to $10^4$
  per second per emitter, as illustrated in Fig. \ref{Fig.7}, which provides a 2D replication of Fig. \ref{Fig.6}. The deviation of the retrieved emitter centroid from its true position toward the overlapping region of emitter PSFs arises from elevated Poissonian noise at low photon count rates, short integration times, and small $\tau$-intervals.  The predominant shot noise over the desired emission signal compromises the SNR of the temporal correlation function, which then Fourier transforms to a noisy, inseparable background superimposed with the desired spectral correlation. The frequency-independent background causes an overestimation of the spectral cross-correlation contribution ($A_{\times}$) during fitting, as it is more susceptible to background noise compared to the spectral auto-correlation contribution due to its broader width. This overestimation is proportionally more pronounced for pixels in the non-overlapping region than for pixels in the overlapping region, given the lower base-case value of $A_{\times}$ at those pixels. Consequently, $A_{\times}$ at pixels in the overlapping region are undervalued after normalization, while the amplitudes of the effective PSF at those pixels, $A^2_{\parallel} - \frac{1}{2}A^2_{\times}$, are overvalued. This overestimation in the PSF amplitude explains the consistent direction of the deviation of emitter localization, specifically, towards the region with high PSF overlap. 
  
  An evident reduction of the Poissonian noise and the systematic error of emitter localization is observed by repeating the simulation with a higher photon count rate, as summarized in Table. \ref{table2}. At low photon count rate, a longer integration time and longer $\tau$-interval result in minimal improvement in localization accuracy due to the quadratic scaling of SNR with the photon count rate but only linear scaling with integration time and $\tau$-interval.

\begin{table}[ht]
\caption{Decreased deviations of emitter localization with higher photon count rate and longer time of integration}
\label{table2}
\resizebox{\textwidth}{!}{%
\begin{tabular}{@{}ccccc@{}}
\toprule
\begin{tabular}[c]{@{}c@{}}Photon count rate \\ $(cps)$ \end{tabular} & \begin{tabular}[c]{@{}c@{}}Time of integration\\ $(s)$\end{tabular} & \begin{tabular}[c]{@{}c@{}}$\tau$-interval\\ $(ms)$\end{tabular} & \begin{tabular}[c]{@{}c@{}}Deviation of first \\ emitter position $(nm)$\end{tabular} & \begin{tabular}[c]{@{}c@{}}Deviation of second\\ emitter position $(nm)$\end{tabular} \\ \midrule
$10^4$            & 1                                                                   & 0.001                                                            &       24.1$\pm$1.7                                                                                &                 18.7$\pm$1.4                                                                      \\
$10^4$            & 1                                                                   & 0.1                                                              &              33.1$\pm$2.1                                                                         &                 17.3$\pm$1.6                                                                      \\
$10^4$ & 10 & 0.1 & 30.7$\pm$2.2 & 12.8$\pm$1.7\\
$10^5$            & 1                                                                 & 0.1                                                              &         0.4$\pm$2.1                                                                              &  7.1$\pm$1.5                                                                                     \\ \bottomrule
\end{tabular}%
}
\end{table}

\newpage

\end{document}